\documentclass[preprint,authoryear,12pt]{imsart}

\usepackage{url}
\usepackage{amsfonts,amsmath,amsthm,graphicx,mathrsfs}
\usepackage{cite}

\usepackage{etoolbox}

\usepackage{graphicx}
\startlocaldefs
\theoremstyle{plain}

\newtheorem{thm}{Theorem}

\newtheorem{lemma}{Lemma}
\newtheorem*{lemma*}{Lemma}

\theoremstyle{remark}

\endlocaldefs

\def\1{\mathbf{1}}

\DeclareMathOperator*{\argmin}{arg\,min}

\def\d{\delta}
\def\e{Z}
\def\bb{\boldsymbol{\beta}}
\def\bm{\boldsymbol{\mu}}
\def\Cov{\mathrm{Cov}}
\def\d{\delta}
\def\D{\boldsymbol{\Delta}}
\def\ee{\mathbf{Z}}
\def\F{\mathcal{F}}
\def\G{\mathscr{G}}
\def\I{\mathbf{I}}
\def\k{\kappa}
\def\l{\lambda}
\def\o{\omega}
\def\R{\mathbb{R}}
\def\s{\sigma}
\def\S{\boldsymbol{\Sigma}}
\def\T{\top}
\def\Var{\mathrm{Var}}
\def\X{\mathbf{X}}
\def\Z{\mathbf{Z}}

\begin{document}

\begin{frontmatter}
\title{Nonparametric empirical Bayes and maximum likelihood estimation for high-dimensional data analysis \protect}

\runtitle{Nonparametric empirical Bayes via maximum likelihood}

\begin{aug}
\author{\fnms{Lee H.} \snm{Dicker}\thanksref{t1}
\ead[label=e1]{ldicker@stat.rutgers.edu}}
\author{\fnms{Sihai D.}\snm{Zhao} \ead[label=e2]{sdzhao@illinois.edu}}

\thankstext{t1}{Supported by NSF Grant DMS-1208785}

\runauthor{L.H. Dicker \& S.D. Zhao}

\affiliation{Rutgers University}

\address{Department of Statistics and Biostatistics \\ Rutgers University \\ Piscataway, NJ 08854 \\
\printead{e1}}
\address{Department of Statistics \\
University of Illinois at Urbana-Champaign \\
Champaign, IL 61820 \\
\printead{e2}}
\end{aug}

\begin{abstract}
Nonparametric empirical Bayes methods provide a flexible and attractive approach to high-dimensional data analysis.  One particularly elegant empirical Bayes methodology, involving the  Kiefer-Wolfowitz nonparametric maximum likelihood estimator (NPMLE) for mixture models, has been known for decades.  However, implementation and theoretical analysis of the Kiefer-Wolfowitz NPMLE are notoriously difficult.  A fast algorithm was recently proposed that makes NPMLE-based procedures feasible for use in large-scale problems, but the algorithm calculates only an approximation to the NPMLE. In this paper we make two contributions. First, we provide upper bounds on the convergence rate of the approximate NPMLE's statistical error, which have the same order as the best known bounds for the true NPMLE.  This suggests that the approximate NPMLE is just as effective as the true NPMLE for statistical applications.  Second, we illustrate the promise of NPMLE procedures in a high-dimensional binary classification problem. We propose a new procedure and show that it vastly outperforms existing methods in experiments with simulated data. In real data analyses involving cancer survival and gene expression data, we show that it is very competitive with several recently proposed methods for regularized linear discriminant analysis, another popular approach to high-dimensional classification.
\end{abstract}

\end{frontmatter}
{\allowdisplaybreaks

\section{Introduction}
\label{s:intro}
Nonparametric empirical Bayes methods offer an attractive approach to analyzing high-dimensional data. The main idea is to treat the unknown high-dimensional parameters as if they were random draws from some common distribution and to estimate the distribution nonparametrically from the data. This estimate is then used as the prior in a full Bayesian analysis. Importantly, these methods often perform effectively, even if the high-dimensional parameters are in fact nonrandom. For example, theoretical and empirical work show that nonparametric empirical Bayes methods perform extremely well when used to estimate the mean of a high-dimensional Gaussian random vector in the frequentist setting \cite{zhang2003compound,zhang2005general,brown2008season,brown2009nonparametric,jiang2009general,brown2013poisson}.   One of the keys to understanding these results lies in the close relationship between empirical Bayes methods and the solutions to compound decision problems \cite{zhang2003compound}.  More broadly, many of the attractive properties of nonparametric empirical Bayes methods may be viewed as generalizing those of the well-known James-Stein estimator \cite{james1961estimation}. 

One particularly elegant nonparametric empirical Bayes methodology involves the Kiefer-Wolfowitz nonparametric maximum likelihood estimator (NPMLE) for nonparametric mixture models \cite{kiefer1956consistency}.  The NPMLE approach to nonparametric empirical Bayes is the focus of this paper;  however, other approaches exist, such as those found in \cite{brown2009nonparametric, greenshtein2009application}, which involve nonparametric smoothing.   Suppose that the data consist of observations $X_1,...,X_N \in \R$ and that $X_j = \mu_j + \e_j$ ($1 \leq j \leq N$), where $\e_1,...,\e_N \sim N(0,1)$ and $\mu_1,...,\mu_N \sim F_0$ are all independent, and $F_0$ is some unknown distribution.  Let $\phi$ denote the standard normal density and let $\F$ denote the class of all probability distributions on $\R$.  The NPMLE estimator for $F_0$ is 
\begin{equation}
  \label{e:npmle}
  \hat{F} = \argmin_{F \in \F} -\sum_{j = 1}^N \log\left\{\int \phi(X_j - \mu) \
  dF(\mu)\right\}.
\end{equation}

One advantage to procedures based on the NPMLE is that there are no tuning parameters; by contrast, careful tuning is typically required for nonparametric smoothing methods.  On the other hand,  computation and theoretical analysis of the Kiefer-Wolfowitz NPMLE are notoriously difficult (see, for example, Chapter 33 of \cite{dasgupta2008asymptotic}).

Recently, Koenker and Mizera \cite{koenker2014convex} proposed a new, scalable method for approximately solving \eqref{e:npmle}.  Their algorithm is based on the observation that \eqref{e:npmle} is a (infinite-dimensional) convex optimization problem.  On the other hand, most previous approaches to solving \eqref{e:npmle} emulate more standard procedures for fitting finite mixture models, i.e. the EM-algorithm \cite{laird1978nonparametric,jiang2009general}, which converge very slowly in the NPMLE setting.  In \cite{koenker2014convex}, Koenker and Mizera illustrate the superior empirical performance of their methods through extensive numerical results. However, they provide no theoretical justification for their estimator, which is only an approximation to the true NPMLE, $\hat{F}$.

In this paper we make two contributions.  First, we derive an upper bound on the rate of convergence of Koenker and Mizera's approximate NPMLE to $F_0$. This upper bound has the same order as the best known bounds for $\hat{F}$, provides stronger theoretical support for Koenker and Mizera's work, and suggests that from a statistical perspective, Koenker and Mizera's approximate NPMLE may be just as effective as $\hat{F}$.  Second, we illustrate the promise of NPMLE procedures in a novel application to high-dimensional binary classification problems. We fit two NPMLEs, $\hat{F}^0$ and $\hat{F}^1$, based on the training data for each group --- $X_j$ is taken to be the mean value of the $j$-th feature for each group --- and then implement the Bayes classifier based on these distribution estimates.  We show that this rule vastly outperforms existing methods in experiments with simulated data, where discriminant-based classification rules have been previously advocated.  In real data analyses, where gene expression microarray data is used to classify cancer patients, we show that the proposed method is very competitive with several recently proposed methods for regularized linear discriminant analysis.

The rest of the paper proceeds as follows.  In Section~\ref{s:convex}, we discuss Koenker and Mizera's approximate NPMLE, which is the solution to a finite-dimensional convex optimization problem. Theoretical results for the approximate NPMLE are presented in Section~\ref{s:theory}.  The application of NPMLEs to high-dimensional binary classification is discussed in Section~\ref{s:class}. In Section~\ref{s:sims} we present the results of numerical experiments with simulated data involving the proposed classification rule.  Section~\ref{s:data} contains the results of several real data analyses involving microrarray data. In Section~\ref{s:corr} we discuss some issues related to the analysis of correlated data. A concluding discussion may be found in Section~\ref{s:disc}.  Proofs, more detailed simulation results, and implementations of strategies for handling correlated data are contained in the Supplementary Material.

\section{Approximate NPMLE via convex optimization}
\label{s:convex}

Lindsay \cite{lindsay1983geometry} showed that the NPMLE, $\hat{F}$, exists and is a discrete measure supported on at most $N$ points in the interval $[X^{(1)},X^{(N)}]$, where $X^{(1)} = \min\{X_1,...,X_N\}$ and $X^{(N)} = \max\{X_1,...,X_N\}$.  Thus, solving \eqref{e:npmle} is equivalent to fitting a finite mixture model with $N$ components. However, it is noteworthy that if we restrict our attention in \eqref{e:npmle} to distributions $F$ that are supported on at most $N$ points, i.e. if we attempt to find $\hat{F}$ by fitting a finite mixture model with $N$ components, then the problem is no longer convex [in contrast, recall from Section 1 that the unrestricted problem \eqref{e:npmle} is convex].

Koenker and Mizera, on the other hand, propose a different type of restriction on $F$. For positive integers $K$, define $\hat{\F}_K$ to be the class of probability distributions supported on the $K+1$ equally spaced points $X^{(1)} = \mu_0 < \mu_1 < \cdots < \mu_K = X^{(N)}$.  Notice that $\hat{\F}_K$ is a random collection of probability distributions, because the atoms $\mu_0,...,\mu_K$ are determined by the data $X_1,...,X_N$.  Koenker and Mizera's approximate NPMLE is 
\begin{equation}
 \label{e:npmlek}
  \hat{F}_K = \argmin_{F \in \hat{\F}_K} -\sum_{j = 1}^N \log\left\{\int \phi(X_j - \mu) \
  dF(\mu)\right\}.
\end{equation}
However, the key property of Koenker and Mizera's NPMLE is that the optimization problem \eqref{e:npmlek} is a convex $K$ dimensional problem.  Moreover, $\hat{F}_K$ can be easily found for $N$ in the 10000s and $K$ in the 100s using an R package developed by Koenker, \texttt{REBayes}, and standard convex optimization solvers \cite{rebayes, mosek}.

\section{Theory for the approximate NPMLE}
\label{s:theory}
In general, $\hat{F}_K \neq \hat{F}$.  However, it is plausible that $\hat{F}_K \to \hat{F}$ as $K \to \infty$, and intuitively the accuracy of $\hat{F}_K$ depends on the parameter $K$.  In \cite{koenker2014convex}, Koenker and Mizera suggest taking $K = 300$, which works well in their examples, but no theoretical justification is provided.  In this section, we present theoretical results which suggest that if $K \approx \sqrt{N}$, then $\hat{F}_K$ is just as accurate as $\hat{F}$ for estimating $F_0$.   Thus, one might expect that Koenker and Mizera's recommendation for $K = 300$ may be reasonable for $N$ up to $\approx 90000$.   Before proceeding, we introduce some additional notation.  

For $x \in \R$, define the convolution densities corresponding to $F_0$, $\hat{F}$, $\hat{F}_K$, 
\[
p_0(x) = \int \phi(x - \mu) \ dF_0(\mu), \ \ \hat{p}(x) = \int \phi(x - \mu) \ d\hat{F}(\mu), \ \ \hat{p}_K(x) = \int \phi(x - \mu) \ d\hat{F}_K(\mu),
\]
respectively.  Additionally, for densities $p,q$ on $\R$ (with respect to Lebesgue measure), define their Hellinger distance 
\[
||p^{1/2} - q^{1/2}||_2 = \left[\int \{p(x)^{1/2} - q(x)^{1/2}\}^2 \ dx\right]^{1/2}.
\]
We measure the distance between $\hat{F}_K$ and $F_0$ by the Hellinger distance between the convolution densities, $||\hat{p}_K^{1/2} - p_0^{1/2}||_2$.  This is somewhat common in the analysis of nonparametric mixture models \cite{ghosal2001entropies}.  Moreover, we note that applications commonly depend on $\hat{F}$ or $\hat{F}_K$ only through the corresponding convolution $\hat{p}$ or $\hat{p}_K$ \cite{zhang2005general, brown2009nonparametric, greenshtein2009application, jiang2009general}, which further suggests that the Hellinger metric for convolutions is reasonable.  

Lastly, before stating our main result, if $\{R_N\}$ is a sequence of random variables and $\{a_N\}$ is a sequence of real number, then the notation $R_N = O_P(a_N)$ means that the sequence $\{R_N/a_N\}$ is bounded in probability.  

\begin{thm}
Suppose that $F_0$ has compact support.  Then 
\begin{equation}\label{e:hbd}
||\hat{p}_{\lfloor N^{1/2}\rfloor}^{1/2} - p_0^{1/2}||_2 = O_P\left\{\frac{\log(N)}{N^{1/2}}\right\},
\end{equation}
where $\lfloor N^{1/2}\rfloor$ is the greatest integer less than or equal to $N^{1/2}$.  
\end{thm}

Theorem 1 is proved in Section~S3 of the Supplementary Material.  It implies that, in the Hellinger metric for convolutions, $\hat{F}_{\lfloor N^{1/2}\rfloor}$ converges to $F_0$ at rate $N^{1/2}/\log(N)$ in probability; this is very nearly the standard parametric rate $N^{1/2}$.  In \cite{ghosal2001entropies}, Ghosal and Van der Vaart proved convergence results for the NPMLE, $\hat{F}$.  Theorem 4.1 of \cite{ghosal2001entropies} implies that under the conditions of our Theorem 1, 
\begin{equation}\label{e:hbd0}
||\hat{p}^{1/2} - p_0^{1/2}||_2 = O_P\left\{\frac{\log(N)}{N^{1/2}}\right\}.
\end{equation}
Observe that the upper bounds in \eqref{e:hbd} and \eqref{e:hbd0} are the same.  The results by Ghosal and Van der Vaart are, to our knowledge, the best upper bounds on the convergence rate of $\hat{F}$ in the Hellinger metric for convolutions.  To our knowledge, there are no specialized lower bounds on the convergence rate of $\hat{F}$.  However, given that the gap between the upper bounds \eqref{e:hbd}--\eqref{e:hbd0} and the parametric rate $N^{1/2}$ is only a factor of $\log(N)$, it seems reasonable to conclude that the statistical properties of $\hat{F}_{\lfloor N^{1/2} \rfloor}$ and $\hat{F}$ are nearly indistinguishable.  

\section{High-dimensional binary classification using NPMLE}
\label{s:class}

Because of their computational convenience, Koenker and Mizera's methods may be used in much broader settings (e.g. larger datasets) than previously possible for NPMLE-based methods.  Moreover, our results in Section \ref{s:theory} provide statistical justification for their use.   Here we illustrate the promise of these methods with a novel application to high-dimensional binary classification. 

\subsection{Setting}
\label{s:setting}

Consider a training dataset with $n$ observations, where $Y_i \in \{0,1\}$ denotes the group membership of the $i$-th observation and $\X_i=(X_{i1},\ldots,X_{iN})^{\T} \in \R^N$  is an associated feature vector ($1 \leq i \leq n$). Our objective is to use the dataset $\mathcal{D}=\{(Y_i,\X_i);i=1,\ldots,n\}$, to build a classifier $\d:\mathbb{R}^N\rightarrow\{0,1\}$ that can accurately predict the group membership $Y^{\mathrm{new}}$ of a subsequent observation, based only on the corresponding feature measurements $\X^{\mathrm{new}} = (X_1^{\mathrm{new}},...,X_N^{\mathrm{new}})^{\T}$. In the high-dimensional problems we consider here, $N$ is much larger than $n$, e.g. $N\approx 10000$, $n \approx 50$.

Assume that the $(N+1)$-tuples $(Y^{\mathrm{new}},\X^{\mathrm{new}}),(Y_1,\X_1),...,$ $(Y_n,\X_n)$ are iid and, for $k = 0,1$, that $\X_i|(Y_i = k) \sim G^k$, where $G^k$ is some $N$-dimensional distribution that has density $g^k$ with respect to Lebesgue measure on $\mathbb{R}^N$.  Assume further that $P(Y_i = 1) = \pi$. The performance of a classifier $\d: \R^N \to \{0,1\}$, which may depend on the training data $\mathcal{D}=\{(Y_i,\X_i); i = 1,...,n\}$ but not the test data $(Y^{\mathrm{new}},\X^{\mathrm{new}})$, will be measured by the misclassification rate $R(\d) = P\{\d(\X^{\mathrm{new}}) \neq Y^{\mathrm{new}}\}$. This probability is computed with respect to the joint distribution of $\mathcal{D}$ and $(Y^{\mathrm{new}},\X^{\mathrm{new}})$. The Bayes rule
\begin{equation}\label{bayes}
\d_B(\X^{\mathrm{new}}) = I\left\{\frac{g^0(\X^{\mathrm{new}})}{g^1(\X^{\mathrm{new}})}\cdot \frac{1-\pi}{\pi} < 1\right\}
\end{equation}
minimizes $R(\d)$ and thus is the optimal classifier \cite{devroye1996probabilistic}. In practice, the densities $g^0,g^1$ (and the probability $\pi$) are unknown, and the Bayes rule cannot be implemented.  Instead, a classification rule $\hat{\d}$ must be constructed from the training data $\mathcal{D}$; often $\hat{\d}$ is constructed to mimic $\d_B$. 

\subsection{Linear discriminant methods}
\label{s:lda}

Classifiers based on Fisher's linear discriminant analysis \cite{fisher1936use} are currently widely used.   A motivating assumption for linear discriminant analysis is that the data are Gaussian; in particular, $G^k=N(\bm^k,\S)$ for $k=0,1$, with $\bm^k=(\mu^k_{1},\ldots,\mu^k_{N})^T\in\R^N$.  In this setting, Fisher's linear discriminant rule is the optimal classifier (\ref{bayes}) and takes the form
\begin{equation}\label{LDA}
\d_F(\X^{\mathrm{new}}) = I\left\{\D^T\S^{-1}(\X^{\mathrm{new}} - \bm) > \log\left(\frac{1 -\pi}{\pi}\right)\right\},
\end{equation}
where $\D = \bm^1 - \bm^0$ and $\bm = (\bm^1 + \bm^0)/2$. In practice, $\bm^0$, $\bm^1$, and $\S$ are typically estimated from the training data, are these estimates are used in place of the corresponding quantities in $\d_F$ to obtain an approximation to the Fisher rule. In high-dimensional problems, where $N$ is large compared to $n$, some type of regularization is required to effectively estimate these parameters, in order to avoid overfitting and excessive noise in the data.   Many regularized Fisher discriminant rules have been proposed, e.g. \cite{friedman1989regularized, fan2008high, greenshtein2009application, witten2011penalized, mai2012direct}.  Other methods related to linear discriminant analysis borrow techniques from robust statistics (in addition to using regularization), in order to better handle non-Gaussian data, e.g. CODA by Han et al. \cite{han2013coda}.

One challenge for many regularized linear discriminant methods is that they require carefully choosing a regularization parameter, which is often done using time-consuming cross-validation. Another potential issue lies in the use of regularization itself.  Regularization is closely connected with Bayesian methods and often amounts to (implicitly or explicitly) imposing a prior distribution on the parameters of interest, e.g. $\bm^0$ and $\bm^1$.  However, if $\bm^0$ and $\bm^1$ truly followed some prior distribution (i.e. if they were truly random), then in general $\d_F$ would no longer be the Bayes rule and would no longer provide optimal classification. This suggests that it may be productive to pursue alternative an alternative approach.  


\subsection{The empirical Bayes NPMLE method}
\label{s:method}
We propose an empirical Bayes NPMLE method for high-dimensional binary classification. In deriving the NPMLE rule, we proceed under the assumption that $\X_i|(Y_i = k,\bm^k) \sim N(\bm^k,\I_N)$ ($k = 0,1)$ and that 
\begin{equation}\label{GMLEBa2}
\mu_1^0,...,\mu_N^0 \sim F^0, \ \  \mu^1_1,...,\mu_N^1 \sim F^1
\end{equation} 
are  independent, where $F^0,F^1$ are some unknown probability distributions on $\R$.  The independence assumptions $\Cov(\X_i|Y_i = k,\bm^k) = \I_N$ ($k = 0,1$) and those involving the $\mu_j^k$ may be viewed collectively as a type of ``naive Bayes'' assumption.  Naive Bayes methods have been commonly used and advocated in high-dimensional classification problems \cite{domingos1997optimality, friedman1997bias, bickel2004some}; however, it is of interest to investigate the possibility of relaxing these assumptions (this is discussed further in Section~\ref{s:corr}).  Additionally, we emphasize that while the NPMLE rule is derived under the assumption that $\bm^0,\bm^1$ are random, this assumption does not appear to be necessary for it to perform effectively in high dimensions; indeed, see the simulation results in Section 5, where $\bm^0,\bm^1$ are fixed, and the theoretical results on empirical Bayes methods for estimating the mean of a random vector \cite{zhang2003compound,zhang2005general,brown2008season,brown2009nonparametric,jiang2009general,brown2013poisson}.

To derive the NPMLE classification rule, suppose without loss of generality that the training data $\mathcal{D}$ are ordered such that $Y_1,...,Y_{n_0} = 0$ and $Y_{n_0 + 1},...,Y_n = 1$, and let $n_1 = n - n_0$.  Additionally, let $\bar{X}_j^0 = n_0^{-1}\sum_{i = 1}^{n_0} X_{ij}$ and $\bar{X}_j^k = n_1^{-1}\sum_{i = n_0+1}^n X_{ij}$, $j = 1,...,N$.  For $k = 0,1$ and $j = 1,...,N$, let $F_j^k$ denote the conditional distribution of $\mu_j^k$, given the training data $\mathcal{D}$.  Then
\[
dF_j^k(\mu) \propto \phi\left\{n_k^{1/2}(\bar{X}_j^k -
  \mu)\right\} \ dF^k(\mu) \ \ \ (k = 0,1, \ j = 1,...,N).
\]
Now define the convolution density 
\[
\phi\star F_j^k(x) = \int_{-\infty}^{\infty} \phi(x - \mu) \
dF_j^k(\mu)
 = \frac{\int_{-\infty}^{\infty} \phi(x - \mu) \phi\{n_k^{1/2}(\bar{X}_j^k -
  \mu)\} \ dF^k(\mu)}{\int_{-\infty}^{\infty} \phi\{n_k^{1/2}(\bar{X}_j^k -
  \mu)\} \ dF^k(\mu)},
\]
for $ x\in \R$. The conditional density of $\X^{\mathrm{new}}|(Y^{\mathrm{new}} = k, \ \mathcal{D})$ then equals $\prod_{j = 1}^p \phi\star F_j^k$; it follows that the Bayes rule (\ref{bayes}) becomes
\begin{equation}\label{star}
\d_{\star}(\X^{\mathrm{new}}) = I\left\{\left[\prod_{j= 1}^p \frac{\phi\star
    F_j^0(X_j^{\mathrm{new}})}{\phi\star F_j^1(X_j^{\mathrm{new}})}\right]\cdot\frac{1-\pi}{\pi} < 1\right\},
\end{equation}
rather than the Fisher discriminant rule \eqref{LDA}.

Using the data $\bar{X}_1^0,...,\bar{X}_N^0$ and $\bar{X}_1^1,...,\bar{X}_N^1$, we can now separately estimate the unknown $F^0$ and $F^1$, respectively, using the approximate NPMLE \eqref{e:npmlek} (with $K = \lfloor N^{1/2}\rfloor$, as suggested by Theorem 1); we denote these estimators by $\hat{F}^0$ and $\hat{F}^1$.  For $j = 1,...,N$, $k = 0,1$ and $x \in \R$, define the estimated convolution density
\[
\phi\star\hat{F}_j^k(x) = \frac{\int_{-\infty}^{\infty} \phi(x - \mu) \phi\{n_k^{1/2}(\bar{X}_j^k -
  \mu)\} \ d\hat{F}^k(\mu)}{\int_{-\infty}^{\infty} \phi\{n_k^{1/2}(\bar{X}_j^k -
  \mu)\} \ d\hat{F}^k(\mu)}.
\]
Our new NPMLE-based classifier is therefore defined as
\begin{equation}\label{GMLEB}
\hat{\d}_{\star}(\X) =  I\left\{\left[\prod_{j= 1}^N \frac{\phi\star
    \hat{F}_j^0(X_j)}{\phi\star \hat{F}_j^1(X_j)}\right]\cdot\frac{1- \hat{\pi}}{\hat{\pi}} < 1\right\},
\end{equation}
where $\hat{\pi}$ is an estimate of $\pi$.  In all of the implementations in this paper, we take $\hat{\pi} = 1/2$; as an alternative, it is often reasonable to take $\hat{\pi} = n_1/n$.  

The NPMLE rule $\hat{\d}_{\star}$ accumulates information across the coordinates of
$\X_i$ and is naturally suited to high-dimensional problems.  Roughly
speaking, if $N$ is large, then $\hat{\d}_{\star}(\X)$ is able to learn more information about the
distributions $F^0$ and $F^1$ from the data, which
increases the likelihood of successful classification.  It is important to note that unlike many regularized discriminant rules, $\hat{\d}_{\star}$ does not directly perform feature selection.  On the other hand, the NPMLE estimates $\hat{F}^0$, $\hat{F}^1$ provide a wealth of information on which other feature selection methods could be based.  For instance, a large value of the posterior mean difference
\[
\int \mu \ d\hat{F}_j^1(\mu) - \int \mu \ d\hat{F}_j^0(\mu) = \frac{\int \mu\phi\{n_1^{1/2}(\bar{X}_j^1 - \mu)\} \ d\hat{F}^1(\mu) }{\int \phi\{n_1^{1/2}(\bar{X}_j^1 - \mu)\} \ d\hat{F}^1(\mu) }  - \frac{\int \mu\phi\{n_0^{1/2}(\bar{X}_j^0 - \mu)\} \ d\hat{F}^0(\mu) }{\int \phi\{n_0^{1/2}(\bar{X}_j^0 - \mu)\} \ d\hat{F}^0(\mu) } 
\]
might be suggestive of an important feature.

\section{Numerical experiments with simulated data}
\label{s:sims}

We compared the NPMLE classifier $\hat{\d}_{\star}$ to several other methods for
high-dimensional binary classification in numerical experiments with simulated data. Each of the alternative methods is a regularized Fisher rule, in the sense of Section~\ref{s:lda}.

The simplest classifier we considered will be referred to as simply the naive Bayes (NB) rule, which replace $\D$ in \eqref{LDA} with $\hat{\D} = (\hat{\Delta}_1,...,\hat{\Delta}_N)^{\T} =  \bar{\X}^1 - \bar{\X}^0 \in \R^N$ and $\S$ with $\I_N$. Theoretical performance of the NB rule in high-dimensions has been studied by \cite{bickel2004some} and many others.

Perhaps the most direct competitor to the NPMLE classifier is a method proposed by Greenshtein and Park in \cite{greenshtein2009application}, which we refer to as GP.  In the GP rule, $\S$ in \eqref{LDA} is replaced with $\I_N$ and $\D$ is replaced with an estimate of the conditional expectation $E(\D|\mathcal{D})$, where the coordinates of $\D$, $\Delta_j = \mu_j^1 - \mu_j^0$ ($j = 1,...,N$), are assumed to be independent draws from some unknown distribution $F$. The distribution of $F$ is estimated nonparametrically by kernel smoothing, using the training data $\mathcal{D}$. Two key differences between GP and our NPMLE rule are: (i) our rule aims to approximate the Bayes rule (\ref{star}), while GP targets the Fisher rule (\ref{LDA}); (ii) our rule uses the Kiefer-Wolfowitz NPMLE to estimate the relevant prior distributions, while GP uses kernel smoothing.  

Another method that we considered in the numerical experiments is an independence oracle thresholding classifier. We replace $\S$ in \eqref{LDA} with $\I_N$ and the difference vector $\D$ with
\begin{equation}\label{Dhat}
\hat{\D}_{\l} =
(\hat{\Delta}_1^{\l},...,\hat{\Delta}_N^{\l})^{\T} \in \R^N, 
\end{equation}
where $\hat{\Delta}_j^{\l} = \hat{\Delta}_jI\{|\hat{\Delta}_j| \geq \lambda\}$ ($j= 1,...,N$) and $\l \geq 0$ is chosen to minimize the misclassification rate on the testing data.  We refer to this method as the oracle naive Bayes (oracle NB) method.  This method shares strong similarities with the FAIR classifier proposed by \cite{fan2008high}; indeed, the oracle NB rule may also be viewed as an oracle version of the FAIR classifier.

We simulated data according to $\X_i=\bm^0 I\{Y_i=0\}+\bm^1 I\{Y_i=1\}+\ee_i \in \R^N$ ($i = 1,...,n$),
where $\ee_1,...,\ee_n \sim N(0,\I_N)$ were independent. Throughout, we took $\bm^0 = 0 \in \R^N$. The vector $\bm^1 = \Delta (m^{-1/2},...,m^{-1/2},0,...,0)^{\T} \in \R^N$ was taken so that the first $m$ components were equal to $\Delta m^{-1/2}$ and the remaining components were equal to 0, for various values of $m$ and $\Delta$.  Observe that the $\ell^2$-norm of $\bm^1$ is $||\bm^1|| = \Delta$ and that $\D = \bm^1 - \bm^0 = \bm^1$.   We emphasize that $\bm^0,\bm^1$ were taken to be fixed vectors in these experiments; in particular, this appears to violate the random $\bm^k$ assumption \eqref{GMLEBa2} underlying the NPMLE rule.

We considered $N = 1000, \ 10000$; $m =10,\  100,\ 500,\ 1000$; and $\Delta = 3,\ 6$.  For each setting we trained the classifiers using $n_1=25$ observations with $Y_i=1$ and $n_0=25$ with $Y_i=0$ (so that $n = n_0 + n_1 = 50$) and tested them on 400 new observations (200 generated from group 0 and 200 generated from group 1). To measure the performance of each classifier, we calculated its misclassification rate over all 400 test observations. We then averaged the rates over 100 simulations.  These simulation settings are similar to those considered by Greenshtein and Park in \cite{greenshtein2009application}.

\begin{figure}[h]
  \centering
  \includegraphics[width=\textwidth]{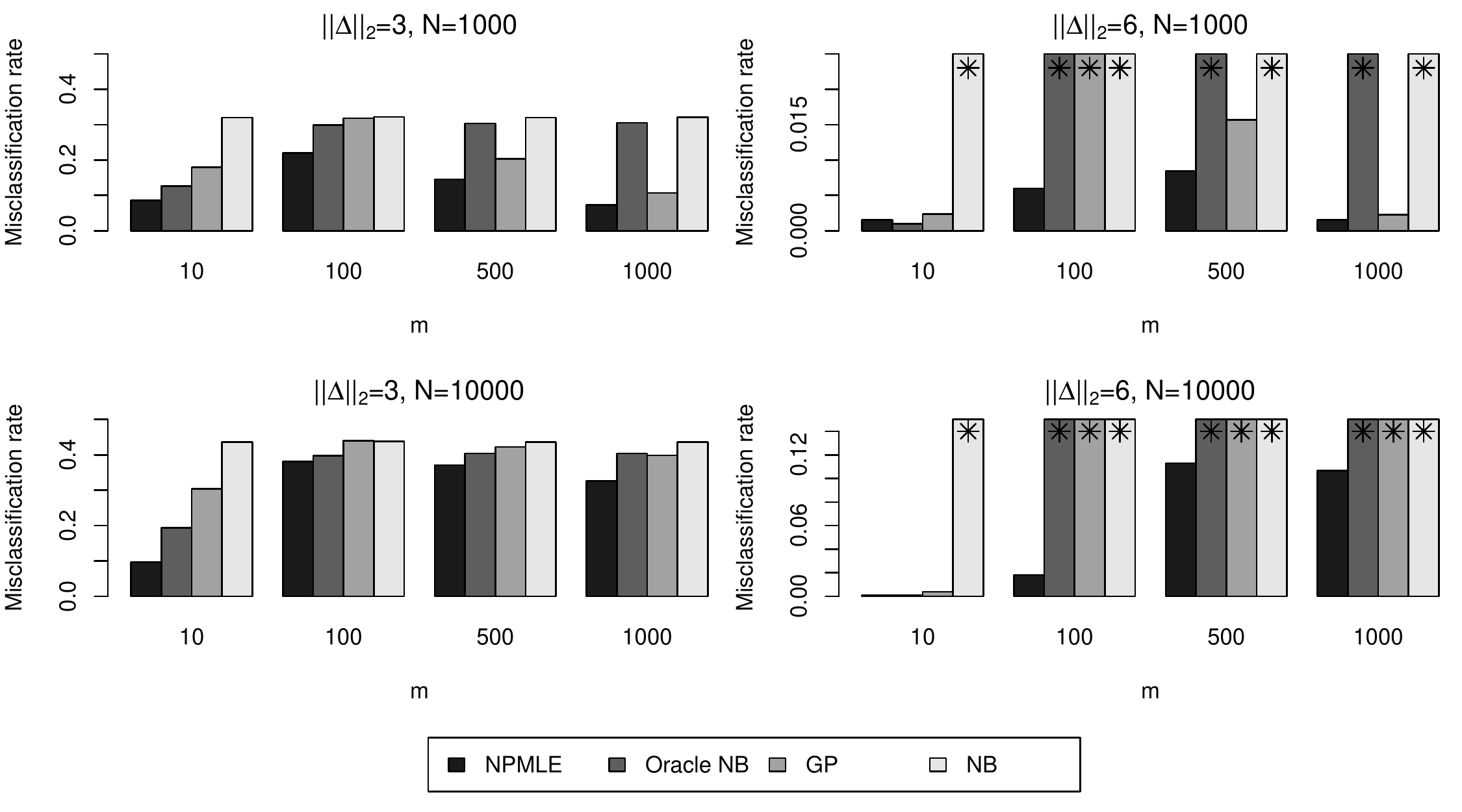}
  \caption{\label{fig:results}Average misclassification rates over 100 simulations. Note the difference in the scale of the y-axes between $\Vert\Delta\Vert_2=3$ and $\Vert\Delta\Vert_2=6$.  Starred columns (``$*$'') indicate that the misclassification rate exceeds the indicated value and has been truncated to fit the plots.}
\end{figure}

Results are presented in Figure~\ref{fig:results}.  The NPMLE procedure gives the lowest misclassification rates in nearly all of the settings. When $N = 10000$ and $(m,\Delta) = (100,6)$, it outperforms all competitors by almost an order of magnitude: it achieves a misclassification error of 0.018, while best error rate of the other classifiers was 0.15, achieved by the oracle NB rule.  More complete results may be found in Section~S1 of the Supplementary Material. The NPMLE rule's performance seems especially impressive when compared with that of the oracle NB rule and the GP rule. Indeed, one might expect that the settings considered here (independent Gaussian errors) would be favorable to oracle NB, which has access to an oracle specifying the optimal thresholding level.  The GP and NPMLE rules both utilize nonparametric empirical Bayes methods; thus, the substantial gains of NPMLE over GP appear noteworthy. These results illustrate the promise of other NPMLE-type methods for related problems in high-dimensional data analysis.

\section{Real data analysis}
\label{s:data}
We also compared our NPMLE rule to the competing methods on three benchmark gene expression datasets. The first comes from a study by Golub et al. \cite{golub1999molecular} of classification of leukemia patients into acute lymphoblastic leukemia (ALL) or acute myeloid leukemia (AML) patients, using expression levels of 7129 genes. The training data consist of 27 subjects with ALL and 11 with AML, while the testing data contain 20 subjects with ALL and 14 with AML.

The second dataset comes from a study by Gordon et al. \cite{gordon2002translation} of classification of lung cancer patients into malignant pleural mesothelioma (MPM) or adenocarcinoma (ADCA) patients, using expression levels of 12533 genes. The training data consist of 16 subjects with MPM and 16 with ADCA, while the testing data contain 15 subjects with MPM and 134 with ADCA.

Finally, the last dataset comes from a study by Shaughnessy et al. 
\cite{shaughnessy2007validated} of classification of myeloma patients
into those surviving for longer or shorter than two years, using the
intensity levels of 54675 probesets. This dataset was used in the
Microarray Quality Control Phase II project
\cite{maqc2010microarray}. We averaged the probeset intensities
corresponding to the same gene symbol, giving 33326 gene expression
levels. The training data contain 288 long-term and 51 short-term
survivors, and the testing data contain 187 long-term and 27
short-term survivors, where long-term survivors lived for more than two years.

We standardized each feature to have sample variance 1. We could not implement the oracle NB because we do no know the true class labels in the testing data. Instead we used the FAIR classifier, proposed by \cite{fan2008high}, which like the oracle NB rule is a thresholding-based classifier. We chose the threshold level $\l$ in (\ref{Dhat}) as in Theorem~4 of \cite{fan2008high}, where the authors derived an expression for the misclassification error, derived the threshold that minimizes this error, and then used the training data to estimate this threshold.

\begin{table*}[t]
  \caption{Misclassification errors on data examples}
\label{tab:results}
\begin{center}
  \begin{tabular}{rccccc}
    \hline
    Dataset & \# test subjects & NPMLE & NB & GP & FAIR\\
    \hline
    Leukemia & 34 & 5 & 6 & 3 & 7 \\
    Lung & 149 & 1 & 1 & 1 & 14 \\
    Myeloma & 214 & 89 & 88 & 108 & 88 \\
    \hline
  \end{tabular}
\end{center}
\end{table*}

The number of misclassification errors in the test data for the various datasets and methods are reported in Table~\ref{tab:results}.  The NPMLE rule was comparable to the other classifiers in the leukemia and lung cancer datasets. In the myeloma data, it performed similar to NB and FAIR.

\section{Strategies for correlated data}
\label{s:corr}
We have so far treated the features as if they were independent, which is unlikely to be true in real applications. While correlation plays an important role in high-dimensional data
analysis, there has been a great deal of theoretical and
empirical work that suggests that treating the features as
independent, even if they are dependent, can be an effective
strategy \cite{domingos1997optimality,friedman1997bias,hand2001idiot,bickel2004some}. We conducted additional simulation experiments, described in Section~S1 of the Supplementary Material, to study the performance of our NPMLE classification rule on correlated Gaussian data. We found that it still outperformed the NB and GP classifiers; its performance was comparable to that of the oracle NB classifier, as well as that of a recently proposed regularized Fisher rule specifically designed to account for correlation \cite{mai2012direct}.

On the other hand, understanding appropriate methods for handling correlation is an important and nontrivial objective. Developing comprehensive NPMLE methods for correlated data is beyond the scope of this paper. However, in Section~S2 of the Supplementary Material we combined our NPMLE rule with a simple generic method for handling correlation in high-dimensional analysis to re-analyze the three gene expression datasets described in Section 6. We screened out highly correlated features, which makes the remaining data appear to be ``more independent'' and more closely approximates the independence assumptions underlying our NPMLE rule. We found that this simple procedure improved performance in the data analysis. Nevertheless it remains of interest to develop NPMLE procedures that explicitly account for dependence.

\section{Discussion}
\label{s:disc}

We believe that the computational convenience of Koenker and Mizera's approximate NPMLE methods will make the use of NPMLE-based methods for nonparametric empirical Bayes far more practical in many applications with high-dimensional data.   In this paper, we derived results that provide theoretical support for Koenker and Mizera's methods, and proposed a novel application to high-dimensional binary classification problems, which illustrates the promise of these methods.   

There are many interesting directions for future research in this area; we mention three below.  While Koenker and Mizera's method greatly simplifies calculation of the (approximate) NPMLE, their implementation relies on a generic convex optimization solver.  Further gains in computational efficiency may be possible by a more detailed analysis of the optimization problem \eqref{e:npmlek}.  Additionally, the location mixture model \eqref{e:npmle} is only the simplest application of nonparametric maximum likelihood estimation. 
The \texttt{REBayes} package implements a number of other algorithms that utilize convex optimization, such as a location-scale mixture model where the NPMLE is used to estimate the distribution of the location and scale parameters \cite{rebayes}.  It may be of interest to further investigate theoretical, computational, and practical aspects of these and other related algorithms.  Finally, we believe that there are a number of ways that the NPMLE classification rule proposed in this paper could potentially be improved. For instance, explicit feature selection could be implemented by thresholding the components of the difference of the posterior mean vectors, as suggested in Section~\ref{s:method}, and methods for handling dependent features need to be further developed.  

\section*{\Large Supplementary material}

\setcounter{section}{0}
    \renewcommand{\thesection}{S\arabic{section}}

\setcounter{table}{0}
    \renewcommand{\thetable}{S\arabic{table}}

\setcounter{figure}{0}
    \renewcommand{\thefigure}{S\arabic{figure}}

\section{Detailed simulation study}
\subsection{Classifiers compared}
In addition to the NPMLE, oracle NB, GP, and NB rules mentioned in the main paper, we also considered the classifier proposed by Mai, Zou, and Yuan \cite{mai2012direct}, which we refer to as MZY. This is a regularized Fisher rule that, unlike
the methods mentioned above, makes a direct effort to account for correlation
between the features.  As with the other methods described above, this rule replaces $\bm$ with $(\bar{\X}^0 + \bar{\X}^1)/2$.  However,
the vector $\S^{-1}\D \in \R^N$ in the Fisher classification rule is replaced with $\hat{\bb}_{\l} \in
\R^N$, which solves the following optimization problem:  
\begin{equation}
\label{lasso}
(\hat{\bb}_{\lambda},  \hat{\beta}_{\lambda_0})
=
\argmin_{(\bb,\beta_0) \in \R^{N+1}} 
\Biggl\{n^{-1}\sum_{i=1}^n (Y_i-\beta_0-\X_i^T\bb)^2 +\l\sum_{j=1}^N |\beta_j|\Biggr\}.
\end{equation}
The parameter $\l \geq 0$ is a tuning parameter.  In the numerical
experiments, we took $\l$ to minimize the misclassification rate on
the testing data. Observe that (\ref{lasso}) is an instance of the lasso problem
\cite{tibshirani1996regression}, and that the MZY classifier
may be viewed as a version of ``lassoed''-discriminant analysis; other
related methods have been proposed by \cite{witten2011penalized} and
\cite{fan2012road}.  In high dimensions, the MZY procedure is
expected to perform well when $\bb = \S^{-1}\D$ is sparse, and not
necessarily when $\D$ itself is sparse; the significance of sparse
$\bb$ has been noted by \cite{mai2012direct} and others
\cite{cai2011direct, fan2012road}.

\subsection{Simulation settings}
We simulated data according to $\X_i=\bm^0 I\{Y_i=0\}+\bm^1 I\{Y_i=1\}+\Z_i \in \R^N$ ($i = 1,...,n$). Throughout, we took $\bm^0 = 0 \in \R^N$. In the main paper we took $\bm^1 = \Delta (m^{-1/2},...,m^{-1/2},0,...,0)^{\T} \in \R^N$ was taken so that the first $m$ components were equal to $\Delta m^{-1/2}$ and the remaining components were equal to 0, for various values of $m$ and $\Delta$. We also let $\Z_1,...,\Z_n \sim N(0,\I_N)$. To assess the robustness of our NPMLE classifier we also $\Z_i$ from other distributions.

Specifically, we simulated heavy-tailed $\Z_i$ such that $\sqrt{3}\Z_i \sim
t_3$ followed a $t$-distribution with 3 degrees of freedom (the $\sqrt{3}$ factor implies that $\Var(\Z_i) = 1$). We did not change $\bm^1$.

We also simulated correlated $\Z_i\sim
N(0,\S)$, where $\S$ was either an AR1 correlation matrix, with the
$jk$th entry equal to $\rho^{\vert j-k\vert}$, or an exchangeable
matrix, where the diagonal entries were equal to one and the
off-diagonal entries were all equal to $\rho$. In this setting we also took $\bm^1 =
2/\sqrt{10} (1,1,1,1,1,1,1,-1,-1,-1,0,...,0)^{\T} \in
\R^N$, where the first 10 coordinates in $\bm^1$ were as specified and
the remaining coordinates were all equal to 0. In the experiments reported
here, we considered $N = 1000, \ 10000$ and
$\rho=0.3,\ 0.5,\ 0.7,\ 0.9$. 

\subsection{Results}
Simulation results for the independent Gaussian errors are reported in the main paper. They are reported again in Figure~\ref{fig:ind}, with the addition of the performance of the MZY classifier. More detailed numerical results are reported in Table~\ref{tab:gauss}. The NPMLE classification rule outperforms MZY, along with every other methods, sometimes by an order of magnitude.

Results for the independent heavy-tailed errors are given in Figure~\ref{fig:t} and Table~\ref{tab:t3}.  Results are
more mixed than for the experiments with independent Gaussian errors.
When $N = 1000$, NPMLE remains the top performer in several
settings. On the other hand, when $N = 10000$, the oracle NB and MZY
methods appear to be dominant; we note, however, that both of these
methods are optimally tuned in these experiments (to minimize the
misclassification rate) and this is generally not possible in
practice.  In the $N = 10000$ settings, the
performance of NPMLE is relatively close to that of GP (slightly outperforming
in some instances, and underperforming in others); furthermore, in all
but the $(m,\Delta) = (500,6)$ and $(10000,6)$ settings, the
performance of NPMLE is fairly close to that of the optimal procedure.

Results for the correlated Gaussian errors are given in Figure~\ref{fig:corr} and Table~\ref{tab:cor}. When $\S$ has an AR1
structure, our NPMLE rule still outperforms the NB and GP
classifiers, though it performs better for smaller $\rho$. In
addition, its performance is comparable to that of the oracle NB and MZY
procedures, even though it does not use sparsity or correlation
information. On the other hand, when $\S$ has an exchangeable
structure, all of the procedures exhibit similar misclassification
rates, with better results for larger $\rho$ and oracle NB having a
slight advantage over the other methods.

\begin{figure}[ht]
  \centering
  \includegraphics[width=\textwidth]{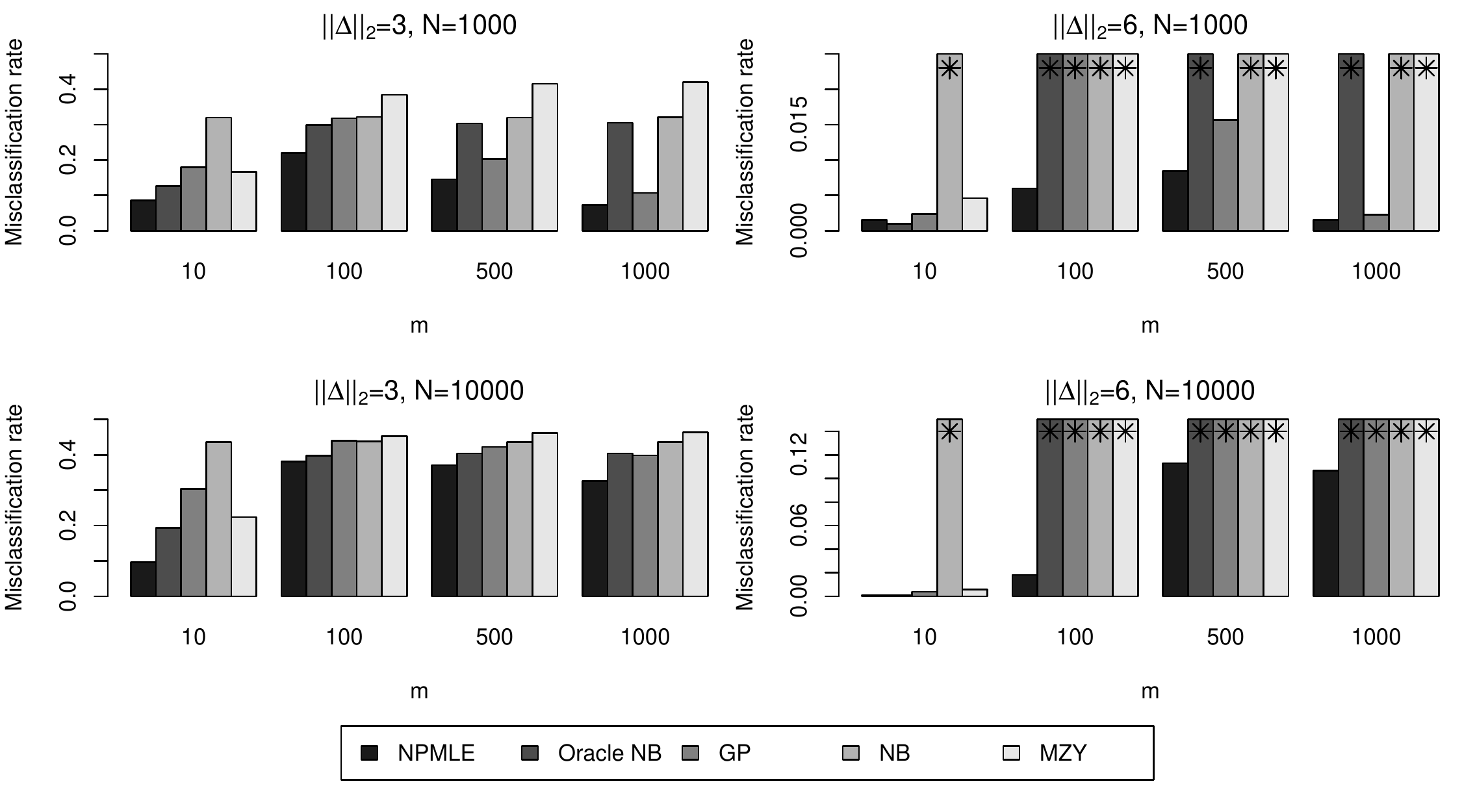}
  \caption{\label{fig:ind}Average misclassification rates over 100
    simulations for independent Gaussian errors. Note the difference
    in the scale of the y-axes between $\Vert\Delta\Vert_2=3$ and
    $\Vert\Delta\Vert_2=6$.  Starred columns (``$*$'') indicate that the misclassification rate exceeds the indicated value and has been truncated to fit the plots.}
\end{figure}

\begin{figure}[ht]
  \centering
  \includegraphics[width=\textwidth]{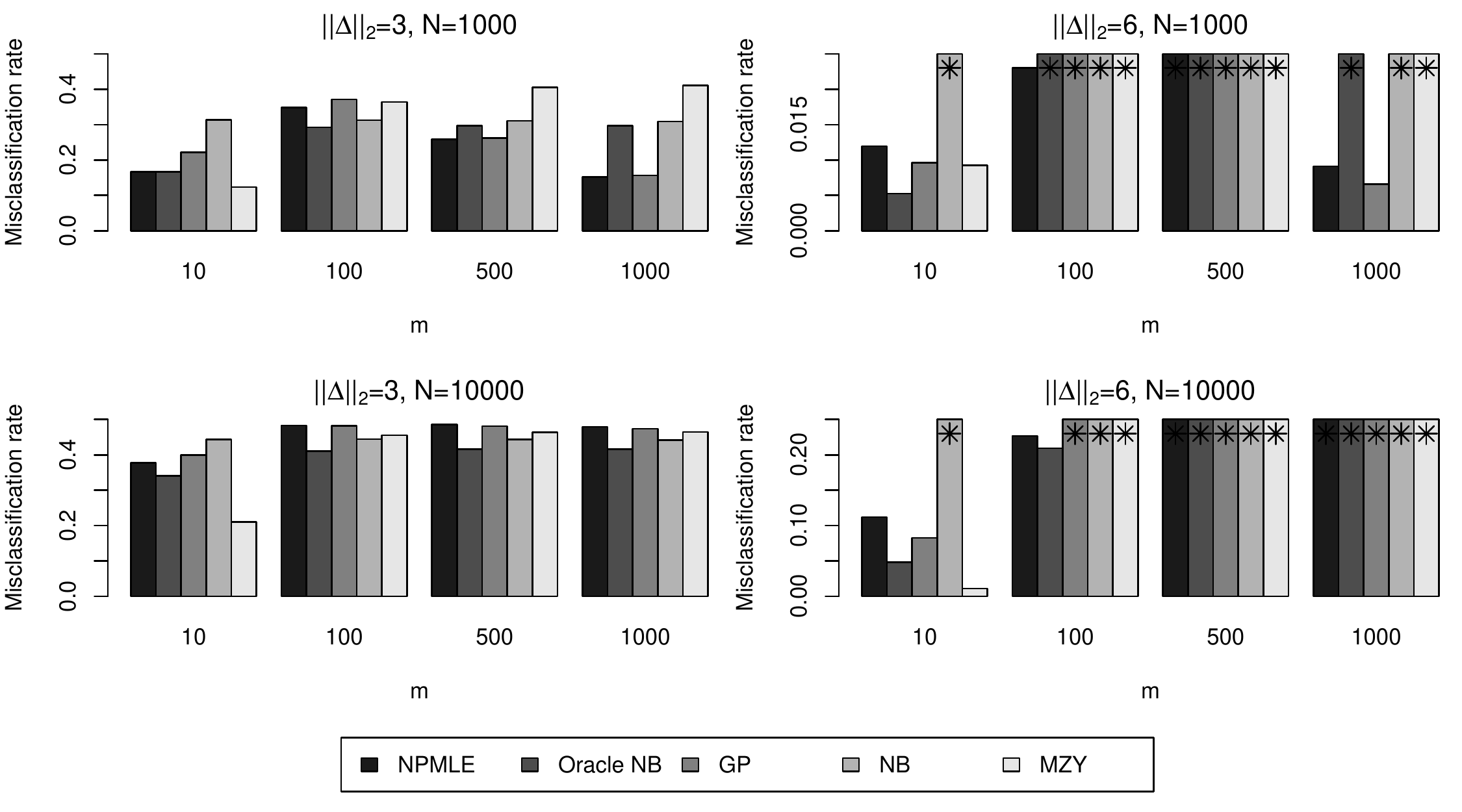}
  \caption{\label{fig:t}Average misclassification rates over 100
    simulations for independent heavy-tailed errors. Note the
    difference in the scale of the y-axes between
    $\Vert\Delta\Vert_2=3$ and $\Vert\Delta\Vert_2=6$.  Starred columns (``$*$'') indicate that the misclassification rate exceeds the indicated value and has been truncated to fit the plots.}
\end{figure}

\begin{figure}[ht]
  \centering
  \includegraphics[width=\textwidth]{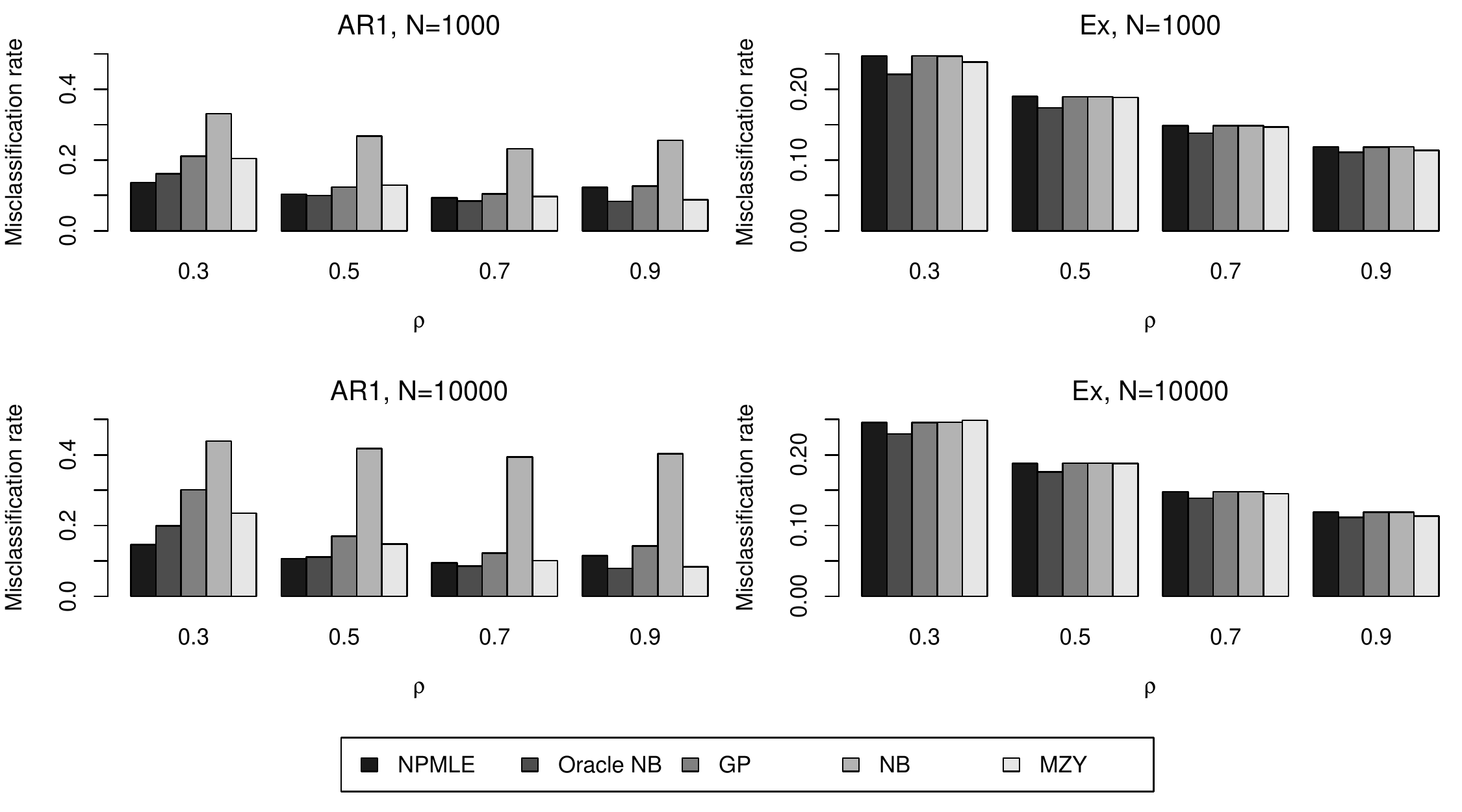}
  \caption{\label{fig:corr}Average misclassification rates over 100 simulations for correlated Gaussian errors. Note the difference in the scale of the y-axes between the AR1 and exchangeable (Ex) simulation settings.}
\end{figure}

\begin{table}[ht]
  \caption{Misclassification rates for data with independent Gaussian errors; ``$*$'' indicates smallest error rate for each setting.}
\label{tab:gauss}
\begin{center}
  \begin{tabular}{r|lllll|lllll}
    & \multicolumn{5}{c|}{$N=1000$} & \multicolumn{5}{|c}{$N=10000$}\\
    \hline
    ($m$,$\Delta$) & NPMLE & NB & GP & Oracle NB & MZY & NPMLE & NB &
    GP & Oracle  NB & MZY \\
    \hline
    (10,3) & 0.085$^*$ & 0.320 & 0.180 & 0.126 & 0.166 & 0.097$^*$ & 0.436 & 0.304 & 0.194 & 0.224 \\ 
    (100,3) & 0.220$^*$ & 0.322 & 0.318 & 0.299 & 0.384 & 0.381$^*$ & 0.438 & 0.440 & 0.398 & 0.453 \\ 
    (500,3) & 0.146$^*$ & 0.320 & 0.203 & 0.303 & 0.415 & 0.371$^*$ & 0.436 & 0.423 & 0.404 & 0.462 \\ 
    (1000,3) & 0.072$^*$ & 0.320 & 0.107 & 0.305 & 0.420 & 0.326$^*$ & 0.436 & 0.398 & 0.404 & 0.463 \\ 
    (10,6) & 0.002 & 0.049 & 0.002 & 0.001$^*$ & 0.005 & 0.001$^*$ & 0.268 & 0.004 & 0.001$^*$ & 0.006 \\ 
    (100,6) & 0.006$^*$ & 0.048 & 0.026 & 0.026 & 0.139 & 0.018$^*$ & 0.266 & 0.158 & 0.150 & 0.250 \\ 
    (500,6) & 0.008$^*$ & 0.047 & 0.016 & 0.042 & 0.286 & 0.113$^*$ & 0.266 & 0.223 & 0.245 & 0.410 \\ 
    (1000,6) & 0.002$^*$ & 0.048 & 0.002$^*$ & 0.043 & 0.320 & 0.107$^*$ & 0.267 &
    0.186 & 0.248 & 0.430 
\end{tabular}
\end{center}
\end{table}

\begin{table}[ht]
 \caption{Misclassification rates for data with independent
   heavy-tailed errors (standardized $t_3$ errors); ``$*$'' indicates smallest error rate for each setting.}
\label{tab:t3}
\begin{center}
  \begin{tabular}{r|lllll|lllll}
    & \multicolumn{5}{c|}{$N=1000$} & \multicolumn{5}{|c}{$N=10000$}\\
    \hline
   ($m$,$\Delta$) & NPMLE & NB & GP & Oracle NB & MZY & NPMLE & NB & GP & Oracle NB & MZY \\  
    \hline
    (10,3) & 0.166 & 0.314 & 0.222 & 0.166 & 0.123$^*$ & 0.378 & 0.444 & 0.399 & 0.340 & 0.210$^*$ \\ 
    (100,3) & 0.348 & 0.313 & 0.371 & 0.293$^*$ & 0.364 & 0.483 & 0.444 & 0.482 & 0.411$^*$ & 0.455 \\ 
    (500,3) & 0.258$^*$ & 0.311 & 0.262 & 0.297 & 0.405 & 0.486 & 0.443 & 0.481 & 0.416$^*$ & 0.464 \\ 
    (1000,3) & 0.152$^*$ & 0.309 & 0.156 & 0.297 & 0.411 & 0.479 & 0.442 & 0.474 & 0.416$^*$ & 0.465 \\ 
    (10,6) & 0.012 & 0.049 & 0.010 & 0.005$^*$ & 0.009 & 0.112 & 0.279 & 0.082 & 0.048 & 0.011$^*$ \\ 
    (100,6) & 0.023$^*$ & 0.049 & 0.045 & 0.029 & 0.114 & 0.227 & 0.281 & 0.302 & 0.209$^*$ & 0.271 \\ 
    (500,6) & 0.031$^*$ & 0.048 & 0.033 & 0.043 & 0.273 & 0.405 & 0.279 & 0.395 & 0.260$^*$ & 0.429 \\ 
    (1000,6) & 0.009 & 0.049 & 0.007$^*$ & 0.045 & 0.307 & 0.403 & 0.278 & 0.379 & 0.263$^*$ & 0.442 \\ 
    \hline
  \end{tabular}
\end{center}
\end{table}

\begin{table}[ht]
  \caption{Misclassification rates for data with correlated Gaussian errors;``$*$'' indicates smallest error rate for each setting.} 
\label{tab:cor}
\begin{center}
  \begin{tabular}{r|lllll|lllll}
    & \multicolumn{5}{c|}{$N=1000$} & \multicolumn{5}{|c}{$N=10000$}\\
    \hline
    $\rho$ & NPMLE & NB & GP & Oracle NB & MZY  & NPMLE & NB & GP & Oracle NB & MZY\\ 
    \hline
    & \multicolumn{5}{c|}{AR1} & \multicolumn{5}{|c}{AR1} \\
    0.3 & 0.136$^*$ & 0.331 & 0.211 & 0.161 & 0.204 & 0.146$^*$ & 0.439 & 0.301 & 0.199 & 0.235 \\ 
    0.5 & 0.103 & 0.268 & 0.124 & 0.100$^*$ & 0.129 & 0.107$^*$ & 0.418 & 0.169 & 0.111 & 0.148 \\ 
    0.7 & 0.093 & 0.232 & 0.104 & 0.084$^*$ & 0.096 & 0.094 & 0.394 & 0.122 & 0.085$^*$ & 0.101 \\ 
    0.9 & 0.123 & 0.256 & 0.126 & 0.083$^*$ & 0.087 & 0.115 & 0.404 & 0.142 & 0.079$^*$ & 0.083 \\ 
    & \multicolumn{5}{c|}{Exchangeable} & \multicolumn{5}{|c}{Exchangeable} \\
    0.3 & 0.247 & 0.247 & 0.247 & 0.221$^*$ & 0.238 & 0.246 & 0.246 & 0.246 & 0.230$^*$ & 0.249 \\ 
    0.5 & 0.190 & 0.189 & 0.189 & 0.174$^*$ & 0.189 & 0.188 & 0.188 & 0.188 & 0.176$^*$ & 0.188 \\ 
    0.7 & 0.149 & 0.149 & 0.149 & 0.138$^*$ & 0.147 & 0.148 & 0.148 & 0.148 & 0.139$^*$ & 0.145 \\ 
    0.9 & 0.119 & 0.119 & 0.118 & 0.111$^*$ & 0.114 & 0.119 & 0.119 & 0.119 & 0.112$^*$ & 0.113 \\ 
    \hline
  \end{tabular}
\end{center}
\end{table}

\section{Real data analysis with correlation screening}
We used the MZY classifier \eqref{lasso} to analyze the real gene expression datasets. Here we selected the tuning parameter $\l$ by three-fold cross-validation refer to as MZY. The results in Table~\ref{tab:results} show that while our NPMLE rule still exhibited superior performance in the leukemia and lung datasets, account for correlation between features seemed to be beneficial in the myeloma dataset.

\begin{table}[ht]
  \caption{Misclassification errors on data examples}
\label{tab:results}
\begin{center}
  \begin{tabular}{rcccccc}
    \hline
    Dataset & \# test subjects & NPMLE & NB & GP & FAIR & MZY\\
    \hline
    Leukemia & 34 & 5 & 6 & 3 & 7 & 8 \\
    Lung & 149 & 1 & 1 & 1 & 14 & 3 \\
    Myeloma & 214 & 89 & 88 & 108 & 88 & 59 \\
    \hline
  \end{tabular}
\end{center}
\end{table}

In order to see if the NPMLE's performance could be easily improved,  we decided to implement it in conjunction with a simple generic method for handling correlation in high-dimensional analysis, inspired by \cite{jiang2004asymptotic} and \cite{dicker2014variance}. For this modified classifier, which is referred to as ``NPMLE+screening'' in Table~\ref{tab:results}, we simply discard highly-correlated features from the dataset before applying NPMLE to the remaining features.  To explain the screening procedure in more detail, results from \cite{jiang2004asymptotic} imply that if the features are independent, then the maximum absolute correlation between features is approximately $2\sqrt{\log(N)/N}$.  Thus, for each pair
of features in the training data with absolute sample correlation greater than
$2\sqrt{\log(N)/N}$, we remove one of them (chosen at random) from the dataset.  (\cite{dicker2014variance} implemented a similar procedure for screening-out correlated predictors in high-dimensional linear models.)

The results for NPMLE+screening appear in the last column of Table~\ref{tab:results}.  It is noteworthy that NPMLE+screening outperforms NPMLE in each of the analyzed datasets except the lung cancer data, where both classifiers make 1 test error. The correlation screening procedure implemented in NPMLE+screening is surely sub-optimal for handling correlation in general. Still, these results are encouraging, because they suggest that simple and effective approaches to improving the performance of NPMLE in the presence of correlation are available.  Further pursuing these ideas is an area of interest for future research.  

\begin{table}[ht]
  \caption{Misclassification errors of NPMLE+screening on data examples}
\label{tab:results_screening}
\begin{center}
  \begin{tabular}{rcccc}
    \hline
    Dataset & \# test subjects & \# features & \# retained & Error\\
    \hline
    Leukemia & 34 & 7129 & 2847 & 4\\
    Lung & 149 & 12533 & 3632 & 1\\
    Myeloma & 214 & 33326 & 4069 & 49\\
    \hline
  \end{tabular}
\end{center}
\end{table}

\section{Proof of Theorem~1}

This section contains a proof of Theorem 1 from the main text.  The
theorem is restated below for ease of reference.  

\begin{thm}
 Suppose that $F_0$ has compact support.  Then 
\begin{equation}\label{e:hbd}
||\hat{p}_{\lfloor N^{1/2}\rfloor}^{1/2} - p_0^{1/2}||_2 = O_P\left\{\frac{\log(N)}{N^{1/2}}\right\},
\end{equation}
where $\lfloor N^{1/2}\rfloor$ is the greatest integer less than or equal to $N^{1/2}$.  
\end{thm}

Before proceeding to the bulk of the proof, we state and prove the
following lemma.

\setcounter{lemma}{0}
    \renewcommand{\thelemma}{S\arabic{lemma}}

\begin{lemma}\label{lemma:hell}
Suppose
that $F_0$ is supported on the compact interval $[a,b]$.  Fix a
positive real number $\D > 0$, and suppose that $\mu = \mu_1 \in \R$ satisfies $a \leq \mu < a + \D$.  Define $K_{\D} = \lfloor(b-a)/\D\rfloor + 1$ and let $\mu_k =
\mu_1 + k\D $ for $k = 2,...,K_{\D}$.  Define $\o_1 = F_0(\mu)$ and
$\o_k = F_0(\mu_k) - F_0(\mu_{k-1})$ for $k = 2,...,K_{\D}$.  Finally,
define the mixture density 
\[
p_{\D,\mu}(x) = \int\phi(x - \mu) \ dF_{\D,\mu}(\mu), 
\]
where 
\[
dF_{\D,\mu}(\mu) = \sum_{k = 1}^{K_{\D}} \o_k\d(\mu - \mu_k) \ d\mu
\]
Then 
\[
||p_{\D,\mu}^{1/2} - p_0^{1/2}||_2 \leq c_{a,b} \D,
\]
where $c_{a,b} \in \R$ is a constant depending only on $a,b$.   
\end{lemma}

{\em Proof of Lemma \ref{lemma:hell}.}  Define $\mu_0 = -\infty$ and let $M = \max\{|a|,|b|\}$. Observe that 
\begin{align*}
\left\{p_{\D,\mu}(x)^{1/2} -
    p_0(x)^{1/2}\right\}^2 & = \frac{\left\{p_{\D,\mu}(x) - p_0(x)\right\}^2}{\left\{p_{\D,\mu}(x)^{1/2} +   p_0(x)^{1/2}\right\}^2} \\
& \leq \frac{1}{p_0(x)}\left\{\int \phi(x - \mu) \ dF_0(\mu) - \int
  \phi(x - \mu) \ dF_{\D,\mu}(\mu)\right\}^2 \\
& =  \frac{1}{p_0(x)}\left[\int \phi(x - \mu) \ dF_0(\mu) \right. \\ &
\qquad \qquad\qquad \left. - \sum_{k = 1}^{K_{\D}}
  \phi(x - \mu_k)\{F_0(\mu_k)- F_0(\mu_{k-1})\}\right]^2\\
& = \frac{1}{p_0(x)}\left[  \sum_{k = 1}^{K_{\D}}\int_{\mu_{k-1}}^{\mu_k} \phi(x - \mu) -
  \phi(x - \mu_k)\ dF_0(\mu)\right]^2 \\
& \leq \frac{1}{\phi(|x| + M)}\left\{\sum_{k= 1}^{K_{\D}} \int_{\mu_{k-1}}^{\mu_k}
  \left|\phi(x - \mu) - \phi(x - \mu_k)\right| \ dF_0(\mu)\right\}^2.
\end{align*}
For any $x \in \R$, we additionally have
\begin{align}\nonumber 
\left\{p_{\D,\mu}(x)^{1/2} -
    p_0(x)^{1/2}\right\}^2 & \leq \frac{1}{2\pi e\phi(|x| +
    M)}\left\{\sum_{k= 1}^{K_{\D}} \int_{\mu_{k-1}}^{\mu_k} 
  \left|\mu - \mu_k\right| \ dF(\mu)\right\}^2 \\ \label{ineq1} & \leq \frac{\D^2}{2\pi e\phi(|x| + M)}.
\end{align}
Furthermore, since $\phi'(|x|)$ is decreasing when $|x| > 1$, if $|x| > M + 1$, then
\begin{align}\nonumber
\left\{p_{\D,\mu}(x)^{1/2} -
    p_0(x)^{1/2}\right\}^2 & \leq \frac{\phi'(|x| - M)^2}{\phi(|x| +
    M)}\left\{\sum_{k= 1}^{K_{\D}} \int_{\mu_{k-1}}^{\mu_k} 
  \left|\mu - \mu_k\right| \ dF(\mu)\right\}^2 \\ \nonumber
& \leq \frac{\phi'(|x| - M)^2}{\phi(|x| + M)}\D^2 \\ \label{ineq2}
& = e^{4a^2}(|x| - a)^2\phi(|x| - 3a)\D^2.
\end{align}
Combining (\ref{ineq1})--(\ref{ineq2}) yields 
\begin{align*}
||p_{\D,\mu}^{1/2} - p_0^{1/2}||_2^2 & \leq \int_{|x| \leq M + 1} \frac{\D^2}{2\pi
  e\phi(|x| + M)}\ dx + \int_{|x| > M +
  1} e^{4a^2}(|x| - a)^2\phi(|x| - 3a)\D^2 \ dx \\
& \leq c_{a,b}^2\D^2.
\end{align*}
The lemma follows.   \hfill $\Box$ 

\vspace{.1in}

Returning to the proof of Theorem 1, we follow techniques very similar
to those found in \cite{wong1995probability}.  Assume that the support of $F_0$
is contained in the closed interval $[a,b]$, for fixed real numbers $a
< b$.  Let $\e> 0$ be a real
number and let $X^{(1)} \leq X^{(2)} \leq \cdots \leq
X^{(N)}$ be the order statistics for the data $X_1,...,X_N$.  Define the events
\begin{align*}
A_K(\e) & = \left\{||\hat{p}_K^{1/2} - p_0^{1/2}||_2 > \e\right\}, \\
B & = \left\{[a,b] \subseteq [X^{(3)},X^{(N-2)}] \subseteq [X^{(1)},X^{(N)}] \subseteq \left[a - \sqrt{8\log(N)},b + \sqrt{8\log(N)}\right]\right\}.
\end{align*}
The probability $P(B)$ is easily bounded.  Indeed, we have
\begin{align*}
P(B) & \geq 1 - P\left\{a < X^{(3)}\right\} - P\left\{X^{(N-2)} <
  b\right\} \\
& \qquad - P\left\{X^{(1)} < a - \sqrt{8\log(N)}\right\} -
P\left\{b + \sqrt{8\log(N)} < X^{(N)}\right\}.
\end{align*}
Clearly, 
\begin{align*}
P\left\{a < X^{(1)}\right\} & \leq \binom{N}{2}P\{a < X_j \}^{N-2} \leq \binom{N}{2}\left\{1 -
  \Phi(a-b)\right\}^{N-2} = \binom{n}{2}\Phi(b-a)^{N-2}\\
P\left\{X^{(N)} <b \right\} & \leq \binom{N}{2}P\{X_j< b\}^{N-2} \leq \binom{n}{2}\Phi(b-a)^{N-2},
\end{align*}
where $\Phi$ is the standard normal CDF. On the other hand, 
\begin{align*}
P\left\{b + \sqrt{8\log(N)} < X^{(N)}\right\} & \leq N P\left\{b +
  \sqrt{8\log(N)} < X_j\right\}  \\
& = NP\left\{
  \sqrt{8\log(N)} < Z_j\right\} \\
& \leq \frac{N}{\sqrt{8\log(N)}}\phi\left\{\sqrt{8\log(N)}\right\} \\
& = \frac{1}{4N^3\sqrt{\pi\log(N)}}
\end{align*}
and, similarly, 
\[
P\left\{X^{(1)} < a - \sqrt{8\log(n)}\right\} \leq  \frac{1}{4n^3\sqrt{\pi\log(n)}}.
\]
We conclude that 
\begin{equation}\label{PB}
P(B) \geq 1 - 2\binom{N}{2}\Phi(b-a)^{N-2} - \frac{1}{2N^3\sqrt{\pi\log(N)}}
\end{equation}
and 
\begin{align}\nonumber 
P\{A_K(\e)\} & = P\{A_K(\e)\cap B\} + P\{A_K(\e) \cap B^c\} \\
& \leq P\{A_K(\e) \cap B\} + 2\binom{N}{2}\Phi(b-a)^{n-2}+ \frac{1}{2N^3\sqrt{\pi\log(N)}}. \label{bd1}
\end{align}
To prove the theorem, we bound $ P\{A_K(\e) \cap B\}$ and choose $K$,
$\e$ appropriately.  

We follow the notation from Lemma \ref{lemma:hell} and, on the event
$B$, consider the
(random) distribution function $F_{\D,\mu}$, where $\mu = X^{(1)} +
(X^{(N)} - X^{(1)})\tilde{k}/K$, $\tilde{k}$ satisfies 
\[
X^{(1)} +
\frac{X^{(N)} - X^{(1)}}{K}(\tilde{k}-1) < a \leq X^{(1)} +
\frac{X^{(N)} - X^{(1)}}{K}\tilde{k},
\]
and $\D = (X^{(N)} - X^{(1)})/K$.  Observe that for constants $c_1 >
0$, 
\begin{align*}
A_K(\e) \cap B &\subseteq \left\{\sup_{||p^{1/2} -
      p_0^{1/2}||_2 > \e, \ p \in \hat{\F}_K} \prod_{j = 1}^N
  \frac{p(X_j)}{p_{\D,\mu}(X_j)} > 1\right\} \cap B \\
& \subseteq \left\{\sup_{||p^{1/2} -
      p_0^{1/2}||_2 > \e, \ p \in \F} \prod_{j = 1}^N
  \frac{p(X_j)}{p_0(X_j)} > e^{-c_1N\e^2} \right\}  \\ & \qquad \qquad \cup \left[\left\{ \prod_{j = 1}^N
  \frac{p_0(X_j)}{p_{\D,\mu}(X_j)} > e^{c_1N\e^2} \right\} \cap B\right].
\end{align*}
Thus,
\[
P\{A_K(\e) \cap B\} \leq P_1 + P_2,
\]
where 
\begin{align*}
P_1 & = P^*\left\{\sup_{||p^{1/2} -
      p_0^{1/2}||_2 > \e, \ p \in \F} \prod_{j = 1}^N
  \frac{p(X_j)}{p_0(X_j)} > e^{-c_1N\e^2} \right\}, \\
P_2 & = P \left[\left\{ \prod_{j = 1}^N
  \frac{p_0(X_j)}{p_{\D,\mu}(X_j)} > e^{c_1N\e^2} \right\} \cap B\right].
\end{align*}
and, as in \cite{wong1995probability}, $P^*$ denotes the
outer-measure corresponding to $P$.  By Theorem 3.1 of
\cite{ghosal2001entropies} and Theorem 1 of
\cite{wong1995probability}, $c_1$ and an additional constant $c_2 >
0$ may be chosen so that if $\e \geq D_{a,b} \log(N)/\sqrt{N}$ for some
sufficiently large constant $D_{a,b} > 0$, which may depend on $(a,b)$,  then
\begin{equation}\label{P1bd}
P_1 \leq 5e^{-c_2N\e^2}.
\end{equation}

It remains to bounds $P_2$.   We have
\begin{equation}\label{empKL}
\left\{ \prod_{j = 1}^N
  \frac{p_0(X_j)}{p_{\D,\mu}(X_j)} > e^{c_1N\e^2} \right\}  =
\left\{\frac{1}{N}\sum_{j = 1}^N
  \log\left\{\frac{p_0(X_j)}{p_{\D,\mu}(X_j)}\right\} > c_1\e^2\right\}
\end{equation}
and our strategy is to bound $P_2$ using the expression (\ref{empKL})
and Markov's inequality.  The challenge is that the summands in the right-hand
side of (\ref{empKL}) are dependent.  To remove this dependence, we
 exploit the fact that the density $p_{\D,\mu}$ depends on
$\{X_1,...,X_N\}$ only through $X^{(1)},X^{(N)}$.  Thus, if we discard
elements of the dataset $\{X_1,...,X_N\}$ and follow the same
procedure for building the density $p_{\D,\mu}$ using the reduced
dataset, the resulting density is the same unless $X^{(1)}$ or
$X^{(N)}$ is among the discarded data.  More specifically, for $j = 1,...,N$, define $p_{\D,\mu}^{(j)}$ to be the density constructed in the same
manner as $p_{\D,\mu}(x)$ using the observed data with the $j$-th
observation removed.   Then $p_{\D,\mu}^{(j)}$ is independent of
$X_j$ and $p_{\D,\mu}^{(j)} = p_{\D,\mu}$ unless $X_j = X^{(1)}$ or
$X^{(n)}$.    Similarly, for distinct $i,j = 1,...,N$, define $p_{\D,\mu}^{(i,j)}$ to be the density constructing in the same
manner as $p_{\D,\mu}(x)$ using the observed data with the $i$-th and $j$-th
observations removed.  

Before proceeding further, we derive a basic inequality for the likelihood ratio
$p_0(x)/p_{\D,\mu}(x)$ [found in (\ref{LRx}) below] that will be used more than once in the
sequel. Let $M =
\max\{|a|,|b|\}$. Notice that 
\begin{align*}
p_{\D,\mu}(x) & = \sum_{k = 1}^{K_{\D}} \phi(x - \mu_k)\{F_0(\mu_k) -
F_0(\mu_{k-1})\}  \\
& \leq \sum_{k = 1}^{K_{\D}} \sup_{\mu_{k-1} \leq \mu \leq \mu_k}
\frac{\phi(x - \mu_k)}{\phi(x - \mu)} \int_{\mu_{k-1}}^{\mu_k} \phi(x
- \mu) \ dF_0(\mu) \\
& = \sum_{k = 1}^{K_{\D}} \sup_{\mu_{k-1} \leq \mu \leq \mu_k}
\exp\left[(\mu_k - \mu) \left\{x - \frac{1}{2}(\mu_k + \mu)\right\}\right] \int_{\mu_{k-1}}^{\mu_k} \phi(x
- \mu) \ dF_0(\mu) \\ 
& \leq  
e^{\D (|x| + M + \D)} \sum_{k = 1}^{K_{\D}} \int_{\mu_{k-1}}^{\mu_k} \phi(x
- \mu) \ dF_0(\mu) \\
& \leq e^{\D (|x| + M + \D)}p_0(x),
\end{align*}
where $\mu_{-1} = a - \d$ for some small $\d > 0$.  Similarly, one can
check that $p_{\D,\mu} \geq e^{-\D (|x| + M + \D)}p_0(x)$ and it
follows that
\begin{equation}\label{LRx}
e^{-\D (|x| + M + \D)} \leq \frac{p_0(x)}{p_{\D,\mu}(x)} \leq e^{\D (|x| + M + \D)}.
\end{equation}

Now we return to analyzing the event (\ref{empKL}).  On the event $B$,
since $p_{\D,\mu} \neq p_{\D,\mu}^{(j)}$ only if $X_j = X^{(1)}$ or $X^{(N)}$,
 (\ref{LRx}) implies that
\[
\frac{1}{N}\left[\sum_{j = 1}^N
  \log\left\{\frac{p_0(X_j)}{p_{\D,\mu}(X_j)}\right\} - \sum_{j = 1}^N
  \log\left\{\frac{p_0(X_j)}{p_{\D,\mu}^{(j)}(X_j)}\right\}\right] \leq \frac{C_{a,b}\log(N)}{NK},
\]
for some constant $C_{a,b} > 0$.  It follows that 
\begin{align*}
\Bigg\{\frac{1}{N}\sum_{j = 1}^N
  \log\left\{\frac{p_0(X_j)}{p_{\D,\mu}(X_j)}\right\} & >
  c_1\e^2\Bigg\} \cap B \\
& \subseteq \left\{\frac{1}{N}\sum_{j = 1}^N
  \log\left\{\frac{p_0(X_j)}{p_{\D,\mu}^{(j)}(X_j)}\right\} >
  c_1\e^2  - \frac{C_{a,b}\log(N)}{NK} \right\}\cap B.
\end{align*}
Now define 
\begin{align*}
\xi^{(j)} & = E\left[
  \left.\log\left\{\frac{p_0(X_j)}{p_{\D,\mu}^{(j)}(X_j)}\right\}
  \right|X_i, \ i \neq j\right], \\
\zeta^{(j)} & = E\left[\left.\left(
  \log\left\{\frac{p_0(X_j)}{p_{\D,\mu}^{(j)}(X_j)}\right\} -\xi^{(j)}
\right)^2\right|X_i, \ i \neq j\right].
\end{align*}
By Lemma \ref{lemma:hell} above and Lemma
4.1 of \cite{ghosal2001entropies}, there are constants
$c_{a,b}^{(1)},c_{a,b}^{(2)}$ depending only on $a,b$, such that on
the event $B$,
\begin{align}\label{infineq1}
\xi^{(j)} & \leq \frac{c_{a,b}^{(1)}}{K^2}\log(N)\log(K),  \\ \label{infineq2}
\zeta^{(j)} & \leq \frac{c_{a,b}^{(2)}}{K^2}\log(N)\{\log(K)\}^2,
\end{align}
whenever $c_{a,b}\sqrt{\log(N)}/K < 1/2$.  We conclude that if
\begin{equation}\label{bige}
c_1\e^2 \geq \frac{C_{a,b}\log(N)}{NK} + \frac{c_{a,b}^{(1)} \log(N)\log(K)}{K^2},
\end{equation}
then 
\[
\left\{\frac{1}{N}\sum_{j = 1}^N
  \log\left\{\frac{p_0(X_j)}{p_{\D,\mu}(X_j)}\right\} >
  c_1\e^2\right\} \cap B \subseteq G \cap B,
\]
where
\[
G  = \left\{ \left[\frac{1}{N}\sum_{j = 1}^N\left(
  \log\left\{\frac{p_0(X_i)}{p_{\D,\mu}^{(j)}(X_j)}\right\} -\xi^{(j)}  \right)\right]^2
> \k^2\right\} 
\]
and
\[
\k^2 = 
  \left\{c_1\e^2 - \frac{C_{a,b}\log(N)}{NK} - \frac{c_{a,b}^{(1)}
      \log(N)\log(K)}{K^2} \right\}^2.
\]

Now we apply Markov's inequality to $P_2$, 
\begin{align*}
P_2 & \leq P(G \cap B) \\
& \leq \frac{1}{\k^2}E \left\{\left[\frac{1}{N}\sum_{j = 1}^N\left(
  \log\left\{\frac{p_0(X_j)}{p_{\D,\mu}^{(j)}(X_j)}\right\} -\xi^{(j)}\right)\right]^2;
   B\right\} \\
& \leq \frac{1}{n^2\k^2}\sum_{j = 1}^N E(\zeta^{(i)};B) + \frac{2}{N^2\k^2}\sum_{1 \leq i < j \leq N}
E\left(W_iW_j; B\right) \\
& \leq \frac{c_{a,b}^{(2)}}{NK^2\k^2}\log(N)\{\log(K)\}^2 + \frac{2}{N^2\k^2}\sum_{1 \leq i < j \leq N}
E\left(W_iW_j; B\right) \\
& \leq \frac{c_{a,b}^{(2)}}{NK^2\k^2}\log(N)\{\log(K)\}^2 + \frac{1}{\k^2}
E\left(W_1W_2; B\right)
\end{align*}
where we have made use of (\ref{infineq2}) in the third inequality and
\[
W_j = 
  \log\left\{\frac{p_0(X_j)}{p_{\D,\mu}^{(j)}(X_j)}\right\} -\xi^{(j)}.
\]
It remains to bound $E(W_1W_2; B)$.  For $1 \leq i,j \leq N$ and $i
\neq j$, define 
\[
W_i^{(j)} = \log\left\{\frac{p_0(X_i)}{p_{\D,\mu}^{(i,j)}(X_i)}\right\} -\xi^{(i,j)},
\]
where 
\[
\xi^{(i,j)} =  E\left[
  \left.\log\left\{\frac{p_0(X_j)}{p_{\D,\mu}^{(i,j)}(X_j)}\right\}
  \right|X_k, \ k \neq i,j\right].
\]
Since $W_1W_2 = W_1^{(2)}W_2^{(1)}$ on the event that $X_1,X_2 \notin
\{X^{(1)},X^{(N)}\}$, it follows that
\begin{align}\nonumber
E(W_1W_2;B) & = E\left\{W_1^{(2)}W_2^{(1)};B\right\} + E\left\{W_1W_2
  - W_1^{(2)}W_2^{(1)};B\right\} \\ \nonumber
& = E\left\{W_1^{(2)}W_2^{(1)};B\right\} \\ \nonumber
& \qquad + 2\sum_{1 \leq k < l \leq N} E\left\{W_1W_2
  - W_1^{(2)}W_2^{(1)};B, \ (X_1,X_2) = (X^{(k)},X^{(l)})\right\} \\ \nonumber
& = E\left\{W_1^{(2)}W_2^{(1)};B\right\} \\ \nonumber
& \qquad + 2\sum_{2 \leq k < l \leq N-1} E\left\{W_1W_2
  - W_1^{(2)}W_2^{(1)};B, \ (X_1,X_2) = (X^{(k)},X^{(l)})\right\}\\
\nonumber 
& \qquad + 2 \sum_{k = 2}^{N-1} E\left\{W_1W_2
  - W_1^{(2)}W_2^{(1)};B, \ (X_1,X_2) = (X^{(1)},X^{(k)})\right\} \\ \nonumber
& \qquad + 2\sum_{k = 2}^{N-1} E\left\{W_1W_2
  - W_1^{(2)}W_2^{(1)};B, \ (X_1,X_2) = (X^{(k)},X^{(n)})\right\} \\ \nonumber
& \qquad + 2E\left\{W_1W_2
  - W_1^{(2)}W_2^{(1)};B, \ (X_1,X_2) = (X^{(1)},X^{(n)})\right\} \\ \label{Wdecomp}
& = E\left\{W_1^{(2)}W_2^{(1)};B\right\} \\ \nonumber
& \qquad + 2 \sum_{k = 2}^{N-1} E\left\{W_1W_2
  - W_1^{(2)}W_2^{(1)};B, \ (X_1,X_2) = (X^{(1)},X^{(k)})\right\} \\ \nonumber
& \qquad + 2\sum_{k = 2}^{N-1} E\left\{W_1W_2
  - W_1^{(2)}W_2^{(1)};B, \ (X_1,X_2) = (X^{(k)},X^{(n)})\right\} \\ \nonumber
& \qquad + 2E\left\{W_1W_2
  - W_1^{(2)}W_2^{(1)};B, \ (X_1,X_2) = (X^{(1)},X^{(n)})\right\}. 
\end{align}
Now we bound the terms in (\ref{Wdecomp}) separately. 
Let $X^{(1)}_{1,2} = \min\{X_3,X_4,...,X_N\}$ and let $X^{(N)}_{1,2} =
\max\{X_3,X_4,...,X_N\}$.  Define the event 
\[
B_{1,2} = \left\{[a,b]
\subseteq [X_{1,2}^{(1)},X_{1,2}^{(N)}] \subseteq \left[a -
\sqrt{8\log(N)},b+\sqrt{8\log(N)}\right]\right\}
\]
and the $\s$-field $\G_{1,2} = \s(X_3,...,X_N)$.  Then
\begin{align*}
\left|E\left\{W_1^{(2)}W_2^{(1)}; B\right\}\right| & =
\left|E\left[E\left\{\left.W_1^{(2)}W_2^{(1)}I(B)\right|\G_{1,2}\right\};B_{1,2}\right]\right| \\
& = 
\Big|E\left[E\left\{\left.W_1^{(2)}W_2^{(1)}\right|\G_{1,2}\right\}; B_{1,2}\right] \\
& \qquad - E\left[E\left\{\left.W_1^{(2)}W_2^{(1)}I(
  B^c)\right|\G_{1,2}\right\}; B_{1,2}\right]\Big| \\
& = \left|E\left[E\left\{\left.W_1^{(2)}W_2^{(1)}I(
  B^c)\right|\G_{1,2}\right\}; B_{1,2}\right]\right| \\
& \leq \left|E\left[E\left\{\left.(W_1^{(2)}W_2^{(1)})^2\right|\G_{1,2}\right\}^{1/2}  P(
  B^c|\G_{1,2})^{1/2}; B_{1,2}\right]\right| \\
& \leq \frac{c_{a,b}^{(2)} \log(N)\{\log(K)\}^2}{K^2}E\left[P(
  B^c|\G_{1,2})^{1/2};B_{1,2}\right] \\
& \leq \frac{c_{a,b}^{(2)} \log(N)\{\log(K)\}^2}{K^2}P(B^c \cap
B_{1,2})^{1/2} \\
& \leq \frac{c_{a,b}^{(2)} \log(N)\{\log(K)\}^2}{K^2}P(B^c) \\
& \leq \frac{C_{a,b}^{(2)} \log(N)\{\log(K)\}^2}{N^{3/2}K^2}
\end{align*}
for some constant $C_{a,b}^{(2)}$.  By (\ref{LRx}),  on the event $B$, 
\[
|W_j|, \ |W_i^{(j)}| \leq 4\frac{\left\{M + \sqrt{8\log(N)}\right\}^2}{K}
\]
Thus, 
\[
E\left\{W_1W_2
  - W_1^{(2)}W_2^{(1)};B, \ (X_1,X_2) = (X^{(k)},X^{(l)})\right\} 
\leq \frac{64}{N(N-1)}\frac{\left\{M + \sqrt{8\log(N)}\right\}^4}{K^2}
\]
We conclude that there is a constant $\tilde{C}_{a,b}$ such that 
\[
|E(W_1W_2;B)| \leq \tilde{C}_{a,b}\frac{\log(N)\left[\log(N) + \{\log(K)\}^2\right]}{NK^2}.
\]
Putting everything together, we have
\[
P_2 \leq \frac{c_{a,b}^{(2)} + \tilde{C}_{a,b}}{NK^2\k^2}\log(N)\{\log(K)\}^2 + \frac{\tilde{C}_{a,b}}{NK^2\k^2}\log(N)^2.
\]
Thus, if (\ref{bige}) holds, then 
\[
P(A_K(\e) \cap B) \leq  5e^{-c_2N\e^2} + \frac{c_{a,b}^{(2)} + \tilde{C}_{a,b}}{NK^2\k^2}\log(N)\{\log(K)\}^2 + \frac{\tilde{C}_{a,b}}{NK^2\k^2}\log(N)^2
\]
Taking $K = \lfloor N^{1/2}\rfloor$ and 
\[
\e = M_0\frac{\log(N)}{N^{1/2}}
\]
for a large constant $M_0 > 0$ yields
\[
P\left\{||\hat{p}_{\lfloor N^{1/2}\rfloor}^{1/2} - p_0^{1/2}||_2 >
  M_0\frac{\log(N)}{N^{1/2}}\right\}  = O\left\{\frac{1}{\log(N)}\right\}.
\]
Theorem 1 follows.

}
\end{document}